\pdfoutput=1
\documentclass[sigconf]{acmart}

\AtBeginDocument{%
  \providecommand\BibTeX{{%
    \normalfont B\kern-0.5em{\scshape i\kern-0.25em b}\kern-0.8em\TeX}}}

\acmPrice{15.00}
\acmISBN{978-1-45
03-XXXX-X/18/06}

\usepackage{threeparttable}
\usepackage{tabu}                     
\usepackage{multirow}                 
\usepackage{multicol}                 
\usepackage{multirow}                
\usepackage{float}                    
\usepackage{makecell}                 
\usepackage{booktabs}                 

\usepackage{graphicx}
\usepackage{float}
\usepackage{subfigure}
\usepackage[ruled,linesnumbered]{algorithm2e}
\begin{document}

\title{AdaptDHM: Adaptive Distribution Hierarchical Model for Multi-Domain CTR Prediction} %

\author{Jinyun Li, Huiwen Zheng, Yuanlin Liu, Minfang Lu, Lixia Wu, Haoyuan Hu}
\affiliation{%
  \institution{Cainiao Network}
  \country{Hangzhou, China}}
\email{{lijinyun.ljy, xiyou.zhw, yuanlin.lyl, luminfang.lmf, wallace.wulx, haoyuan.huhy}@alibaba-inc.com}

\renewcommand{\shortauthors}{Li et al.}

\begin{abstract}
Large-scale commercial platforms usually involve numerous business domains for diverse business strategies and expect their recommendation systems to provide click-through rate (CTR) predictions for multiple domains simultaneously. Existing promising and widely-used multi-domain models discover domain relationships by explicitly constructing domain-specific networks, but the computation and memory boost significantly with the increase of domains.
To reduce computational complexity, manually grouping domains with particular business strategies is common in industrial applications.
However, this pre-defined data partitioning way heavily relies on prior knowledge, and it may neglect the underlying data distribution of each domain, hence limiting the model's representation capability.
Regarding the above issues, we propose an elegant and flexible multi-distribution modeling paradigm, named \textbf{A}daptive \textbf{D}istribution \textbf{H}ierarchical \textbf{M}odel (AdaptDHM), which is an end-to-end optimization hierarchical structure consisting of a clustering process and classification process. Specifically, we design a distribution adaptation module with a customized dynamic routing mechanism. Instead of introducing prior knowledge for pre-defined data allocation, this routing algorithm adaptively provides a distribution coefficient for each sample to determine which cluster it belongs to. Each cluster corresponds to a particular distribution so that the model can sufficiently capture the commonalities and distinctions between these distinct clusters. Extensive experiments on both public and large-scale Alibaba industrial datasets verify the effectiveness and efficiency of AdaptDHM: Our model achieves impressive prediction accuracy and its time cost during the training stage is more than 50\% less than that of other models.

\end{abstract}


\maketitle

\section{Introduction}
Click-through rate (CTR) prediction is crucial in online recommendation systems, as its performance affects the user experience and is closely tied to the platform's revenue. Although tremendous progress \cite{dnnDeepFM,dnnDIEN,dnnDSIN} has been made in the CTR prediction task, most methods focus on single-domain prediction and suppose that data comes from a homogeneous domain.
Here, a \textbf{\textit{business domain}} is defined as a specific spot where items are displayed to users on applications or websites \cite{STAR}.
In the real-world situation, it is common for large-scale commercial companies (e.g., Alibaba, Amazon) to involve multiple business domains. Taking Alibaba's Taobao app as an example, one of the world's leading online shopping applications, ranking models have been widely applied in hundreds of domains, such as \textit{Homepage}, \textit{Shopping Cart page}, etc.

Recently, some state-of-the-art work has achieved impressive success in this multi-domain recommendation task \cite{HMoE,SAML,STAR}. Inspired by multi-task learning (MTL) \cite{mtlPLE,mtlCrossStitch}, they discover domain relationships by explicitly constructing domain-specific networks. Notably, such a modeling approach is impractical to implement with the rapidly growing business domains due to the significant computation and memory costs. It is common to manually group domains with particular business strategies in industrial applications to reduce computational complexity. 
However, this pre-defined data partitioning way heavily relies on prior knowledge, and it may neglect the underlying data distribution of each domain, hence limiting the model's representation capability.
In addition, it is not flexible for custom pattern modeling. Specifically, distributions under different perspectives are distinct: the data distribution varies among different user groups (e.g., male vs. female users, active vs. cold users), different categories of recommended items (e.g., electronic products vs. cosmetics), etc.

Regarding the above problems, the key to this task is to develop an effective multi-distribution learning strategy to facilitate model learning. An intuitive approach is to introduce a clustering process into the deep CTR model so that it can group similar samples into the same representation space. Then the model can learn the distinctions and commonalities between these spaces comprehensively. For this purpose, we propose an elegant and flexible multi-distribution modeling paradigm, named Adaptive Distribution Hierarchical Model (AdaptDHM), which is an end-to-end optimization hierarchical (multi-level) structure consisting of a clustering process and classification process. A distribution adaptation module with a customized dynamic routing mechanism is designed to allocate multi-source samples into distinct clusters adaptively. Each cluster corresponds to a particular distribution. At last, the model can further learn commonalities and distinctions between these clusters effectively when distribution consistency can be guaranteed in each cluster.

To summarize, the main contributions of this work are as follows:
\begin{itemize}
  \item To the best of our knowledge, this is the first attempt to propose an adaptive distribution modeling paradigm in multi-domain CTR prediction instead of introducing prior knowledge for manually pre-defined domain-aware representation allocation. Also, this modeling framework has strong flexibility to tackle the diverse distribution modeling requirements.
  \item We design a novel distribution adaptation module with a customized dynamic routing mechanism. This routing algorithm provides a distribution coefficient for each instance to determine which distribution it belongs to.
  \item We conduct extensive experiments on both public and Alibaba industrial datasets, and results show that our proposed model outperforms previous work while consuming less memory space and training time.
\end{itemize}

\section{Related Work}

\noindent \textbf{Multi-Domain Learning.} 
Some pioneering work \cite{HMoE,M2M,SAML} formulated multi-domain learning as a special form of multi-task learning (MTL) \cite{mtlMMoE,mtlPLE}, in which common knowledge was shared in the bottom layer and task-aware knowledge was learned in separate branch networks. STAR \cite{STAR} proposed a star topology consisting of shared centered parameters and domain-specific parameters for explicitly exploiting domain relationships.

\noindent \textbf{Dynamic Routing.} 
Our method is inspired by a clustering-like approach called dynamic routing mechanism, firstly proposed by Hinton \cite{sabour2017dynamic} in his capsule network to learn the part-whole relationships iteratively.
Later, Hinton et al. improved the dynamic routing procedure through Expectation-Maximization (EM) algorithm \cite{sabour2018matrix}. Recently, dynamic routing mechanism has been applied in many fields \cite{zhao2018investigating,lei2019multi, kwabena2020exploring}. MIND \citep{li2019multi} utilized dynamic routing in recommender systems to cluster users' historical behaviors and obtained diverse interest representation. 

\section{Proposed Method}

\subsection{Problem Formalization}
We formulate \textit{Multi-Domain CTR Prediction} as a problem that given a set of business domains $\{D_m\}^M_{m=1}$ with a common feature space $\mathcal{X}$ and label space $\mathcal{Y}$. The goal is to construct an unified CTR prediction function $\digamma$: $\mathcal{X}$ $\rightarrow$ $\mathcal{Y}$, which can accurately provide CTR prediction results for $M$ domains simultaneously ($D_1, D_2, ..., D_M$). Unlike previous work, in this paper, we match $M$ given domains into several potential inherent domains $\{D_k\}^K_{k=1}$ with distinct distributions by defining a projection function $\mathcal{G}$: $\mathcal{X}$ $\rightarrow$ $\mathcal{\theta}$. Then, given the input data $\mathcal{X}$ and potential domain $\mathcal{\theta}$, we  attempt to find the function $\digamma$: $\mathcal{X, \theta}$ $\rightarrow$ $\mathcal{Y}$, to provide CTR predictions.

\begin{figure}[t]
  \centering
  \includegraphics[width=0.8\linewidth]{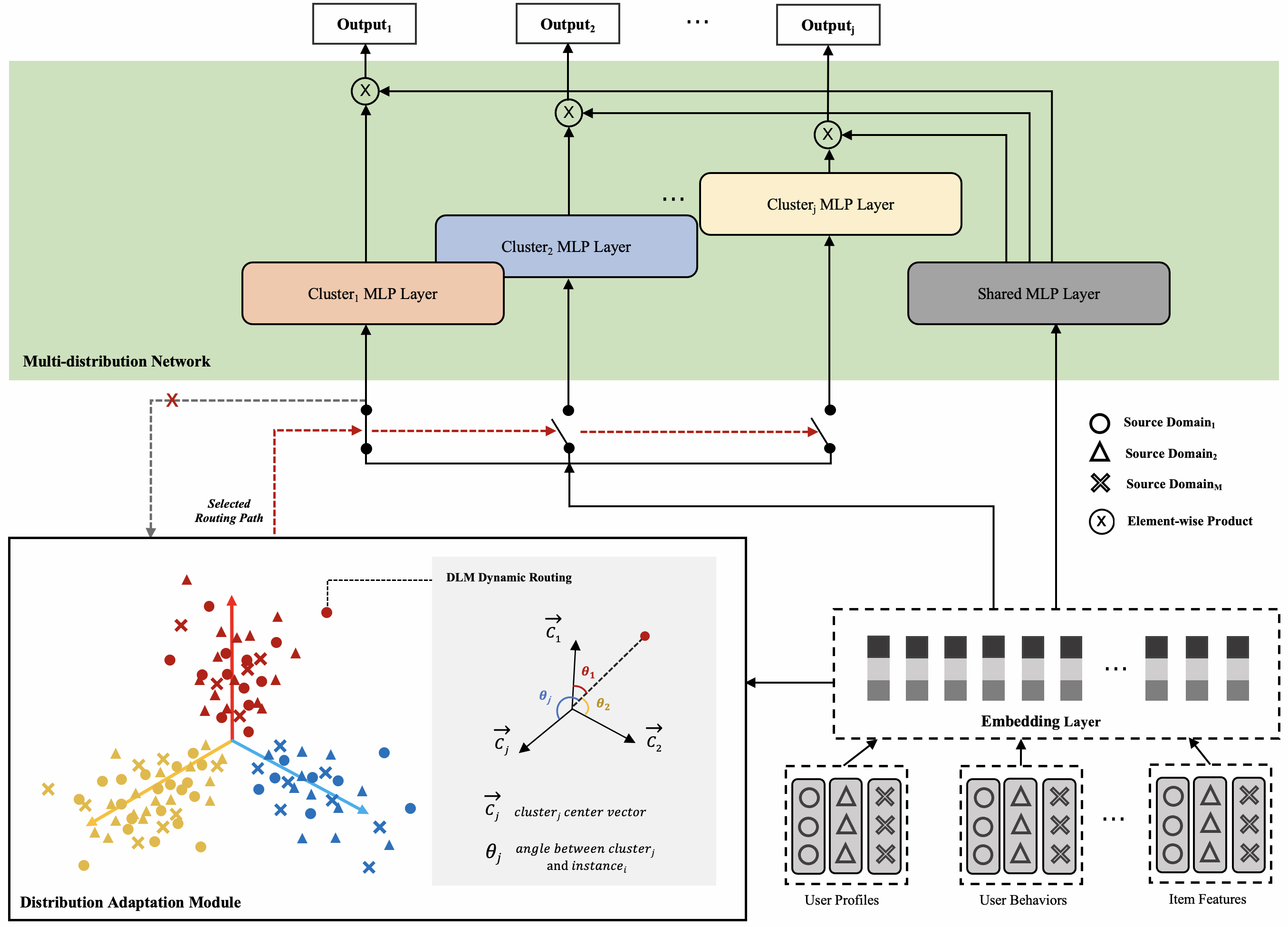}
  \caption{The framework of AdaptDHM. 
  The distribution adaptation module determines which cluster the instance is routed to based on a customized dynamic routing mechanism. Then, the multi-distribution networks capture cluster relationships. The grey line with the red cross prohibits the back-propagation of the gradients.}
  \label{AdaptDHM}
\end{figure}

\subsection{Architecture Overview}
We propose a novel multi-distribution modeling paradigm called AdaptDHM. As shown in Figure \ref{AdaptDHM}, it is an end-to-end optimization hierarchical (multi-level) structure consisting of a clustering process and classification process. 
Specifically, a batch of multi-source instances with various features (e,g., user profile, user's historical behaviors, item features) are fed into the embedding layer to be represented as low-dimensional dense vectors. 

Next, the core component of AdaptDHM, distribution adaptation module, is designed to determine which potential distribution $C_{j}$ ($j \in \{1,...,K \}$) the instance belongs to based on a customized dynamic routing mechanism. 
Then, a standard multi-branch network structure is adopted to learn the cluster commonalities and distinctions from a shared multi-layer perceptron (MLP) layer and cluster-specific layers, respectively. Concretely, the shared MLP layer updates its parameters with all instances, while the specific-cluster network updates its parameters only associated with its corresponding instances. 
Note that the dynamic routing process iterates without introducing backpropagation. At last, the estimated results of the recommended items are produced based on the combination of the shared and cluster-specific parameters.

\subsection{Distribution Adaptation Module}
\noindent \textbf{Input-aware Distribution Modeling.} 
Our model has strong flexibility to accommodate various modeling requirements by feeding different representation vectors. For example, we can construct the underlying user-wise distribution when inputting user-related features. On the other hand, it learns instance-wise distribution when entire features are fed.

\noindent \textbf{DLM Dynamic Routing.}
In distribution learning, we expect small inner-cluster distances while large outer-cluster distances. Inspired by capsule network \cite{sabour2017dynamic} which incorporates clustering and EM algorithms into deep learning, we introduce this idea in our distribution learning. The cosine similarity metric is applied in our work to guarantee that the cosine of the angle between two vectors projected in the same distribution space is small but large in different distribution spaces.
Specifically, we propose the concept of \textbf{\textit{distribution center}} in our designed customized routing algorithm named Distribution Learning Module (DLM) dynamic routing, in which the similarity is calculated between data and each distribution center rather than the measurements between data.
The overall procedure of DLM dynamic routing is shown in Algorithm \ref{alg:algorithm1}.
Given current batch step $b(b\in 1,...,B)$, a batch of embedding vectors $\vec{\mathbf{e}_i}$ ($i \in \{1,...,n\}$), number of cluster $K$, cluster center vectors $\vec{\mathbf{c}_j}$ ($j \in \{1,...,K\}$), iteration times $I$, it returns distribution coefficient $r_{ij}$ which determines the probability of input vector $\vec{\mathbf{e}_i}$ belonging to a particular cluster $j$.

At start of training, we initialize cluster center vectors $\vec{c_j}^{0}$ in form of unit vectors obeying gaussian distribution $\mathcal{N}(0,{\sigma}^2)$. In each iteration of the current batch, we first compute similarity scores $s_{ij}$ of each embedding vector $\vec{\mathbf{e}_i}$ and cluster center vectors $\vec{\mathbf{c}_{j}}^{b}$ based on a cosine similarity metric
\begin{equation}
    s_{ij} 
    = |\vec{\mathbf{c}_{j}}^{b}||\vec{\textbf{e}_i}| \cos \theta
    = \vec{\mathbf{c}_{j}}^{b} \cdot \vec{\textbf{e}_i} 
    ,
\end{equation}
where $\theta$ is the angle between two vectors. Next, the distribution coefficient is obtained by performing softmax of similarly scores
\begin{equation}
    r_{ij}=softmax(s_{ij})=\frac{exp(s_{ij})}{ {\textstyle \sum_{j=1}^{K}}exp(s_{ij}) },
\end{equation}
where the sum of distribution coefficients equals one. Then, the cluster center vector $\vec{\mathbf{c}_{j}}^{b}$ is updated through the weighted sum of distribution coefficients, and turned into unit vector through L2-normalization, denoted as
\begin{equation}
    \vec{\mathbf{c}_{j}}^{b} =  norm_2(\sum_{i=1}^{n}{r_{ij}\vec{\textbf{e}_i}}) ,
\end{equation}
\begin{equation}
\label{eq-norm}
    norm_2(\vec{\mathbf{c}_{j}}^{b})=\frac{1}{\parallel \vec{\mathbf{c}_{j}}^{b} \parallel}_2 \vec{\mathbf{c}_{j}}^{b},
{\parallel \vec{\mathbf{c}_{j}}^{b} \parallel}_2= ({\textstyle \sum_{j=1 }^{K}|\vec{\mathbf{c}_{j}}^{b}|^{2}})^{\frac{1}{2} }.
\end{equation}

In each batch training, all cluster center vectors are inherited the values form the previous batch. To make the value of $\vec{\mathbf{c}_{j}}^{b}$ update stably during the training stage, we apply an exponentially weighted moving average (EWMA) \cite{EWMA} method in our algorithm
\begin{equation}
\vec{\mathbf{c}_{j}}^{b} =  norm_2( \beta * \vec{\mathbf{c}_{j}}^{b-1} + (1-\beta) * \vec{\mathbf{c}_{j}}^{b}), 
\end{equation}
where $\beta$ represents the update rate, set to 0.9 \cite{hu2020design}. 
The larger the value, the smoother the update. 
In this way, the cluster center vector $\vec{\mathbf{c}_{j}}^{b}$ can be smoothly updated, accounting for the value at the current batch and the information from the previous batch.

After training, the cluster center vectors are inherited and used during the inference phase so that the module can guide samples to flow to the cluster with the most similar distribution.

\begin{algorithm}[t]
  \caption{DLM Dynamic Routing.}
  \label{alg:algorithm1}
  \KwIn{current batch step $b(b\in 1,...,B)$, representation vector ${\vec{\mathbf{e}_i}}(i \in 1,...,n$),
  number of clusters $K$,
  iteration times $I$, initialized unit vectors of cluster centers $\vec{\mathbf{c}_j}^{0} \sim \mathcal{N}(0,{\sigma}^2),j \in 1,...,K$ }
  \KwOut{distribution coefficient $r_{ij}$}
  \BlankLine
    Inherited cluster centers $\vec{\mathbf{c}_j}^{b} \gets \vec{\mathbf{c}_j}^{b-1}$\;
    \For {$t$ iterations}{
        for all input vector $i$: $s_{ij} \gets \vec{\mathbf{c}_{j}}^{b} \cdot \vec{\textbf{e}_i}$\;
        for all input vector $i$: $r_{ij} \gets softmax(s_{ij})$\;
        for all cluster center $j$: $\vec{\mathbf{c}_{j}}^{b} \gets  norm_2( \sum_{i=1}^{n}{r_{ij}\vec{\textbf{e}_i}}) $\;
  }
  $\vec{\mathbf{c}_{j}}^{b} \gets  norm_2( \beta * \vec{\mathbf{c}_{j}}^{b-1} + (1-\beta) * \vec{\mathbf{c}_{j}}^{b})$ \;
  \textbf{Return: }$r_{ij}$
\end{algorithm}
\subsection{Multi-distribution Network}
There are many possible ways \cite{mtlMMoE,mtlPLE,STAR} we can adopt to exploit cluster relationships. This paper applies a common multi-branch structure consisting of a shared MLP layer and $K$ cluster-specific MLP layers.
Concretely, parameters in the shared MLP layer are updated by all samples, but cluster-specific MLP layers' parameters are updated only by their corresponding samples. 
We denote the weights of the shared MLP layer and cluster-specific ones as $\mathbf{W}_s$, and $\mathbf{W}_{j}$, respectively.
The final weight $\mathbf{W}_m$ for the $j$-th cluster is
\begin{equation}
\begin{split}
    \mathbf{W}_{m}=\mathbf{W}_{s} \otimes \mathbf{W}_{j},
\end{split}
\end{equation}
where $\otimes$ represents the element-wise product. The prediction result of instance $i$ is produced through a sigmoid function 
\begin{equation}
    \hat{y}_i=sigmoid((\mathbf{W}_{m})^{T}x_{i}).
\end{equation}

The objective function applied in our model is the cross entropy loss function, defined as:
\begin{equation}
\mathcal{L}=-\frac{1}{N}\sum_{i=1}^{n}{( y_i\log(\hat{y}_i)+(1-y_i)\log(1-\hat{y}_i))},
\end{equation}
where $y_i$ is the ground truth of instance $x_i$.

\section{Experiments}

\subsection{Experimental Setup}
\subsubsection{Datasets.}
We conduct extensive experiments on public and industrial datasets. Both of them are gathered from real-world traffic logs of the recommender system.

\noindent \textbf{Public Dataset.} Ali-CCP \cite{Ali-CCP} is a public dataset with training and testing set sizes of over 42.3 million and 43 million, respectively. Since there are too few sources of samples, just three source domains, we attach more features (domain indicator, user gender, user city) to partition samples, resulting to 33 domains.

\noindent \textbf{Industrial Dataset.} 
We collect 6 billion samples from Alibaba online advertising system on 10 business domains. These business domains are manually pre-defined based on prior business knowledge, and each involves tens of sub-domains. We apportion the samples into training and test sets with 60\% and 40\%, respectively, along the time sequence.

\subsubsection{Competitors.}
We use DNN \cite{DNN} as the backbone of all models.
\noindent \textbf{Shared Bottom.} We adapt Shared Bottom model \cite{mtl1997} for multi-domain learning, where the number of task towers is set to the number of domains ($M$) and sharing the embedding layer.\\
\noindent \textbf{PLE.} PLE \cite{mtlPLE} fuses representation learned from shared experts and domain-specific experts.\\
\noindent \textbf{STAR.} 
STAR \cite{STAR} uses a star topology consisting of a centered network and $M$ domain-specific networks. For each domain, a unified model is obtained by element-wise multiplying the weights of the shared network and those of the domain-specific network.

\subsubsection{Implementation Details.}
For a fair comparison, each MLP in all models has the same depth of hidden layers with 5 layers (512-256-128-64-32). The activation function is set to Relu. For the dynamic routing process, we set iteration to 3 \cite{sabour2017dynamic}, update rate $\beta$ to 0.9. Based on the model performance, cluster number is set to 3 (Production) and 9 (Public). For the optimizer, we apply the Adam \cite{Adam} with a batch size of 2048 (Production) and 16384 (Public). The learning rate is set to 1e-3.
All experiments are implemented on a distributed TensorFlow-based framework \cite{xdl}.

\subsubsection{Evaluation Metric.}
In public dataset, we use AUC as evaluate metric. As for industrial dataset, we employ a variant of AUC, named Group AUC (GAUC) \cite{dnnDIN,STAR}, since it is more applicable to comparing online performance in a recommender system. GAUC averags the AUC of different sessions under corresponding impressions. The calculation formula is as follows:
\begin{equation}
    \text{GAUC}=\frac{ {\textstyle \sum_{i=1}^{n}} \left ( \#impression_i \times \text{AUC}_i \right ) }{ {\textstyle \sum_{i=1}^{n}} \#impression_i},
\end{equation}
where $n$ is the number of sessions, $\#impression_i$ and $\text{AUC}_i$ are the number of impressions and AUC with the $i$-th session, respectively.

\begin{table}[t]
  \centering
  \caption{Overall performance comparisons on public dataset and industrial dataset.}
\resizebox{\linewidth}{!}{
    \begin{tabular}{lcc}
    \toprule
\multirow{2}{*}{\textbf{Model}} & \textbf{Public Dataset} & \textbf{Production Dataset}  \\
\cmidrule(lr){2-2}\cmidrule(lr){3-3}
      & AUC & GAUC \\
\hline
    DNN   & 0.6162  & 0.6276    \\
    \quad+Shared Bottom    & 0.5948           & 0.6287  \\
    \quad+PLE             & 0.6152           & 0.6277  \\
    \quad+STAR             & 0.6159           & 0.6286  \\
    \cmidrule(lr){0-0}
    \quad+AdaptDHM               & \textbf{0.6179}  & \textbf{0.6299}\\
    \bottomrule
    \end{tabular}
}
  \label{tab:auc_all}
\end{table}
\begin{table}[t]
  \centering
  \caption{ Single domain Comparison on industrial dataset.}
\resizebox{\linewidth}{!}{
    \begin{tabular}{lllllll}
    \toprule
Domain & DNN & Shared Bottom & PLE   & STAR  & AdaptDHM \\
\hline
    \#1   & 0.6595  & 0.6557  & 0.6623          & 0.6679   & \textbf{0.6838 } \\
    \#2   & 0.6494  & 0.6487  & 0.6480          & 0.6493   & \textbf{0.6521 } \\
    \#3   & 0.6159  & 0.6170  & 0.6152          & 0.6165   & \textbf{0.6178 }  \\
    \#4   & 0.6148  & 0.6163  & 0.6156          & 0.6157   & \textbf{0.6181}  \\
    \#5   & 0.6559  & 0.6567  & 0.6558  & \textbf{0.6571 } &         0.6569  \\
    \#6   & 0.6208  & 0.6235  & 0.6228          & 0.6228   & \textbf{0.6238 } \\
    \#7   & 0.6488  & 0.6497  & 0.6487  & \textbf{0.6503 } &         0.6502  \\
    \#8   & 0.6503  & 0.6569  & 0.6599          & 0.6584   & \textbf{0.6602 } \\
    \#9   & 0.6556  & 0.6563  & 0.6551          & 0.6567   & \textbf{0.6570 } \\
    \#10  & 0.6313  & 0.6328  & 0.6319          & 0.6331   & \textbf{0.6337 } \\
    \hline
    overall GAUC & 0.6276  & 0.6287  & 0.6277  & 0.6286  & \textbf{0.6299}  \\
    \bottomrule
    \end{tabular}
    }
  \label{tab:auc_domain}
\end{table}
\subsection{Effectiveness Verification}
From results in Table \ref{tab:auc_all} and Table \ref{tab:auc_domain}, we have several important observations: (1) AdaptDHM achieves best performance over 1‰ improvements on both public and industrial datasets. Note that 0.1\% AUC gain is regarded as a great advance for the CTR task; (2) DNN performs better than other multi-domain models on the public dataset but worse on the industrial dataset. The main difference between these two datasets is that 
data in the industrial dataset is partitioned manually based on prior solid business knowledge. However, the public one is divided by some randomly-picked domain-aware features. It indicates that explicit modeling way heavily relies on valid data partitioning for domain-specific learning. (3) AdaptDHM shows impressive generalization capability on 10 business domains and outperforms other models in most domains.

\subsection{Hyper-parameters Influence}
\begin{figure}[t]
  \centering
  \includegraphics[width=0.9\linewidth]{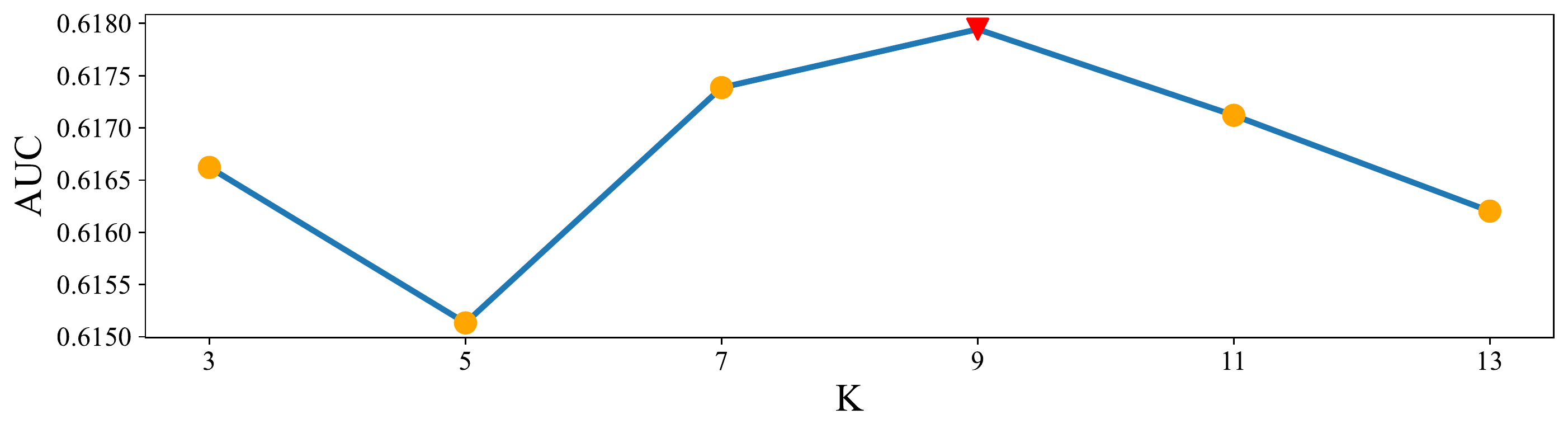}
  \caption{Effect of different cluster number $K$.}
  \label{fig:K}
\end{figure}

We analyze the effect of cluster number $K$ on AdaptDHM's prediction performance. As depicted in Figure \ref{fig:K}, AdaptDHM achieves the best AUC on the public dataset when $K$ equals 9. The performance becomes worse with the growth of $K$. It suggests that that too small or too large $K$ cannot yield the best performance.

\subsection{Efficiency Analysis}

\textbf{Memory and Computation Complexity.} Our model is parameter-efficient owing to the elegant framework. We denote domain-specific MLP parameters as $P_{mlp}$ (about millions of parameters). The number of domains, the number of shared MLP layer (or called shared expert in PLE) and the number of clusters refer to $M$, $S$ and $K$, respectively. The memory and computation cost of each model can be illustrated as Table \ref{tab:par_num1}. 
\begin{table}[ht]
  \centering
  \caption{ Parameters of different models.}
\resizebox{\linewidth}{!}{
    \begin{threeparttable}
    \begin{tabular}{llll}
    \toprule
    SharedBottom & PLE & STAR & AdaptDHM \\
\hline
     $P_{mlp}\times M$ & $P_{mlp}\times (M+S)$ & $P_{mlp}\times (M+1)$ & $P_{mlp}\times (K+1) $ \\
    \hline
    \end{tabular}
    
    \begin{tablenotes}
    \footnotesize
    \item{*}$S \geq 1, and \  K<<M$
    \end{tablenotes}
    \end{threeparttable}
}

  \label{tab:par_num1}
  
\end{table}

\begin{figure}[t]
  \centering
  \includegraphics[width=0.9\linewidth]{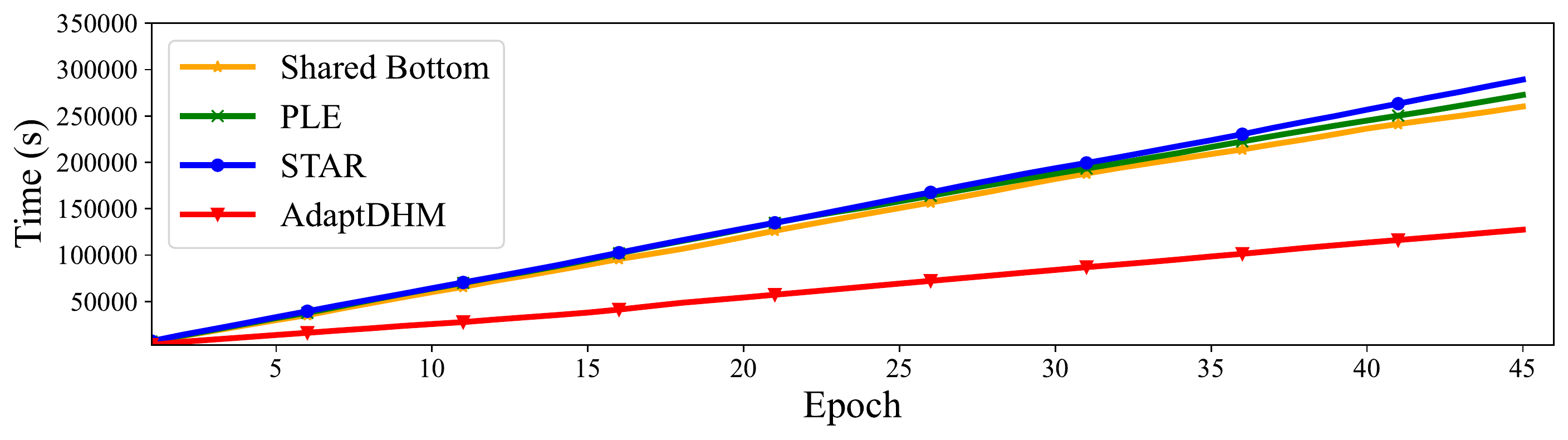}
  \caption{Time consumption with the increase of epochs.}
  \label{fig: times}
\end{figure}

\noindent \textbf{Severing Efficiency.}
Figure \ref{fig: times} shows that AdaptDHM has obvious superiority over others on time cost, taking 50\% less time than others during the training stage, which contributes to more frequent model optimization and online deliveries.

\section{Conclusion}
This paper has proposed an elegant and flexible multi-distribution modeling paradigm, AdaptDHM, to tackle the multi-domain CTR prediction task. Unlike other explicit modeling methods, we have designed a novel distribution adaptation module with a customized dynamic routing mechanism to allocate samples into distinct clusters to facilitate model learning. We have conducted extensive experiments on both public and large-scale Alibaba industrial datasets to verify the effectiveness and efficiency of AdaptDHM. 

\bibliographystyle{plainnat}
\bibliography{cainiao}

\end{document}